\documentclass[aps,prl,twocolumn,groupedaddress,showpacs]{revtex4}

\usepackage{amsmath}
\usepackage{amssymb}
\usepackage{graphicx}% Include figure files

\begin{document}

%\preprint{}

\title{Colloidal crystal growth at externally imposed nucleation clusters}
\author{Sven van Teeffelen}
\email[]{teeffelen@thphy.uni-duesseldorf.de}
\author{Christos N. Likos}
\author{Hartmut L\"owen}
 \affiliation{Institut f\"ur Theoretische Physik II, Weiche Materie,
Heinrich-Heine-Universit\"at D\"usseldorf, 
D-40225 D\"usseldorf, Germany}
\date{\today}

\begin{abstract}
  We study the conditions under which and how an imposed cluster of
  fixed colloidal particles at prescribed positions triggers crystal
  nucleation from a metastable colloidal fluid. Dynamical density
  functional theory of freezing and Brownian dynamics simulations are
  applied to a two-dimensional colloidal system with dipolar
  interactions. The externally imposed nucleation clusters involve
  colloidal particles either on a rhombic lattice or along two linear
  arrays separated by a gap. Crystal growth occurs after the
  peaks of the nucleation cluster have first relaxed to a cutout of
  the stable bulk crystal.
\end{abstract}
%Colloids, 82.70.Dd
%Nucleation in crystal growth, 81.10.Aj
%Liquid/solid transitions, 64.70.Dv
%Crystal growth, 81.10.-h
\maketitle
While important steps towards a quantitative understanding of
homogeneous crystal nucleation out of the melt have been made in the
past decade (for recent reviews, see~\cite{Auer:05,Ruckenstein:05}),
work on the molecular principles of heterogeneous nucleation is still
at its infancy~\cite{Sear:07,Granasy:04,Granasy:06}. Colloidal
suspensions have served as excellent model systems for nucleation,
since the crystallization process is typically much slower than in
their molecular counterparts and the critical nucleus can be detected
in real space~\cite{Gasser:01}. By using external fields, e.g.,
optical tweezers, it is possible to fix a cluster of colloidal
particles and watch directly its impact on the rest of the colloidal
suspension. If the crystal phase is slightly off-coexistence and the
fluid is stable, it is possible to generate crystalline layers around
such a cluster~\cite{Heni:00,Vossen:04,Cacciuto:05}.

In this Letter we study crystal growth processes into a metastable
fluid.  A cluster of fixed colloidal particles, which could act as a
seed for heterogeneous crystal nucleation is arranged within the
metastable melt. Whereas in homogeneous nucleation such clusters
spontaneously form by means of thermal fluctuations, here they are
externally imposed. We investigate whether they serve as initiators of
crystal growth processes.  Our motivation for this study is twofold:
first, by imposing a seed cluster one can steer the crystallization
behavior.  Second, offering complex cluster structures could lead to
unexpected dynamical scenarios of crystal growth.

We approach the problem using classical density functional theory
(DFT) of freezing which is a microscopic approach to
crystallization~\cite{Singh:91,Wu:06}. DFT can be extended to describe
dynamics in strongly inhomogeneous Brownian
fluids~\cite{Marconi:99,Dzubiella:03,Archer:04,PRLBagchifootnote}.
Here it is put forward as the first full microscopic approach to the
dynamics of crystallization.  Our DFT results are backed by Brownian
dynamics (BD) computer simulations. In principle, the dynamical DFT is
superior to phase-field crystal theories of
nucleation~\cite{Elder:02}, which operate on more coarse-grained
length and time-scales and need phenomenological mobilities as an
input. Therefore our results provide benchmark data to test the
validity of more approximate theories.

In detail, we study a model for a two-dimensional suspension of
superparamagnetic colloids, exposed to an external magnetic field
which tunes their parallel dipole moments~\cite{Zahn:97,Haghgooie:05}.
By using additional fields, such as optical tweezers, certain
particles can be fixed in the
suspension~\cite{Koppl:06,PRLPertsinidis}. We first consider a stable
fluid phase, realized for a weak magnetic field. In this fluid
suspension, colloidal particles are placed by optical tweezers into
prescribed positions forming a cluster.  Then the magnetic field is
suddenly increased rendering the fluid metastable with respect to the
stable hexagonal crystal and the tweezers are released. Two different
kinds of cluster geometries are considered: In the first setup we
study hexagonal clusters that are cut out of a perfect rhombic
lattice while in the second setup two sets of linear crystalline
arrays, separated by a gap, are examined.

As a result, we observe that the kinetic pathway of the system is a
{\it two-stage} dynamical process: first, on a sub-Brownian time
scale, the peak positions of the externally imposed nucleation cluster
relax towards a cutout of the stable bulk crystal. Then, on a Brownian
time scale, there are two further possibilities: either the relaxed
cluster acts as a nucleation seed for further complete crystal growth
or it dies out completely without stimulating further crystallization.
Whether crystal growth occurs or not depends delicately on the
compatibility of the initial cluster geometry with that of the stable
bulk crystal in terms of strain energy.

Our system is characterized by the pairwise interaction potential
$u(r)=u_0/r^3$, where $u_0$ is the interaction strength.  For the
specific realization of two-dimensional paramagnetic colloids of
susceptibility $\chi$ exposed to a perpendicular magnetic field ${\bf
  B}$, we have $u_0 = (\chi {\bf B})^2/2$.  The thermodynamics and
structure depend only on one dimensionless coupling parameter $\Gamma
= u_0 \rho^{3/2}/k_BT$, where $\rho$ is the average one-particle
density and $k_BT$ is the thermal energy.

It has been shown~\cite{Marconi:99,Dzubiella:03,Archer:04} that the
static, classical DFT can be given an extension to dynamics to
describe overdamped, time-dependent, out-of-equilibrium systems in
terms of a deterministic, time-dependent, and ensemble averaged
one-particle density $\rho({\bf r},t)$.  The time evolution of
$\rho({\bf r},t)$ is then governed by the continuity equation
\begin{equation}\label{eq:ddft}
  \frac{\partial \rho({\bf r},t)}{\partial t}
  =\frac{D}{k_BT}\nabla \cdot \left[\rho({\bf r},t)\nabla
\frac{\delta F\left[\rho({\bf r},t)\right]}{\delta \rho({\bf r},t)}\right]\,.
\end{equation}
Here, $D/k_BT$ is the mobility coefficient originating from the
solvent, ignoring hydrodynamic interactions.

The equilibrium phase diagram of the system under study has been
obtained using classical DFT~\cite{Teeffelen:06}, which provides the
intrinsic Helmholtz free energy functional $F[\rho({\bf r})]$, a
unique functional of the static one-particle density $\rho({\bf r})$
of the system. The functional $F[\rho({\bf r})]$ is minimized by the
equilibrium one-particle density, where it takes the value of the
system's intrinsic Helmholtz free energy.  The density functional is
typically split into the ideal gas, an excess, and an external part, $
F\left[\rho({\bf r})\right]= F_{\rm{id}}\left[\rho({\bf r})\right]
+F_{\rm{ex}}\left[\rho({\bf r})\right]+F_{\rm{ext}}\left[\rho({\bf
    r})\right]$.  The ideal part is $F_{\rm{id}}\left[\rho({\bf
    r})\right]= k_B T \int\,\mathrm d {\bf r} \rho({\bf r})\left\{
  \ln\left[\rho({\bf r})\Lambda^2\right]-1\right\}$, with $\Lambda$
denoting the thermal de Broglie wavelength. $F_{\rm{id}}$ is of
completely entropic nature and leads to a simple diffusion term in
Eq.~(\ref{eq:ddft}).  The excess part $F_{\rm ex}$, originating from
the correlations between the particles, is in this paper approximated
by the ansatz of Ramakrishnan and Yussouff to the
DFT~\cite{Ramakrishnan:79}. It is expanded up to second order in terms
of density difference $\Delta\rho=\rho({\bf r})-\rho$ around a
reference fluid, where the fluid density $\rho$ is chosen as the
average density of the inhomogeneous system:
$F_{\rm{ex}}\left[\rho({\bf r})\right]\simeq
F_{\rm{ex}}(\rho)-\frac{1}{2}k_BT\iint\,\mathrm d {\bf r}\mathrm d
{\bf r}^\prime \Delta\rho({\bf r})\Delta\rho({\bf
  r}^\prime)c_0^{(2)}({\bf r}-{\bf r}^\prime;\rho)$. Here
$F_{\rm{ex}}(\rho)$ and $c_0^{(2)}({\bf r};\rho)$ are the excess free
energy and the direct correlation function of the reference fluid of
density $\rho$, respectively. Finally, the external part is simply
given by $F_{\rm{ext}}\left[\rho({\bf r})\right]=\int\,\mathrm d {\bf
  r}\rho({\bf r})V({\bf r})$, where $V({\bf r})$ is the total external
potential.

For both setups under study, the clusters of tagged particles are
first, i.e., for times $t<0$, held fixed in a thermodynamically
stable, equilibrated fluid of density $\rho$ at a coupling constant of
$\Gamma_<=10$, which is well below the freezing transition at $\Gamma
\simeq 35.7$~\cite{SvenPhilMag}, obtained within the theory. For the
equilibration of the fluid, Eq.~(\ref{eq:ddft}) is numerically solved
fixing the tagged particles by deep parabolic external potentials at
the tagged particle positions --- in an experiment this could be
achieved by using optical tweezers~\cite{Koppl:06}. At time $t=0$ we
turn the external pinning potential off and, at the same time,
instantaneously quench the system to a coupling constant
$\Gamma_>=62.5$, which is well above the freezing transition and we
observe the time evolution of the density field for times
$(t/\tau_B)\lesssim 10$, where $\tau_B=(\rho D)^{-1}$ is the Brownian
time scale.  Eq.~(\ref{eq:ddft}) is numerically solved applying a
finite difference method. The dimensions $L_x\times L_y=n_x a \times
n_y(\sqrt{3}/2)a$ of the rectangular periodic box considered are
chosen integer multiples $n_x$, $n_y$ of the lattice spacing
$a=(2/\sqrt{3})^{1/2}\rho^{-1/2}$ of the perfectly ordered hexagonal
crystal.

\begin{figure}
  \includegraphics[width=5cm]{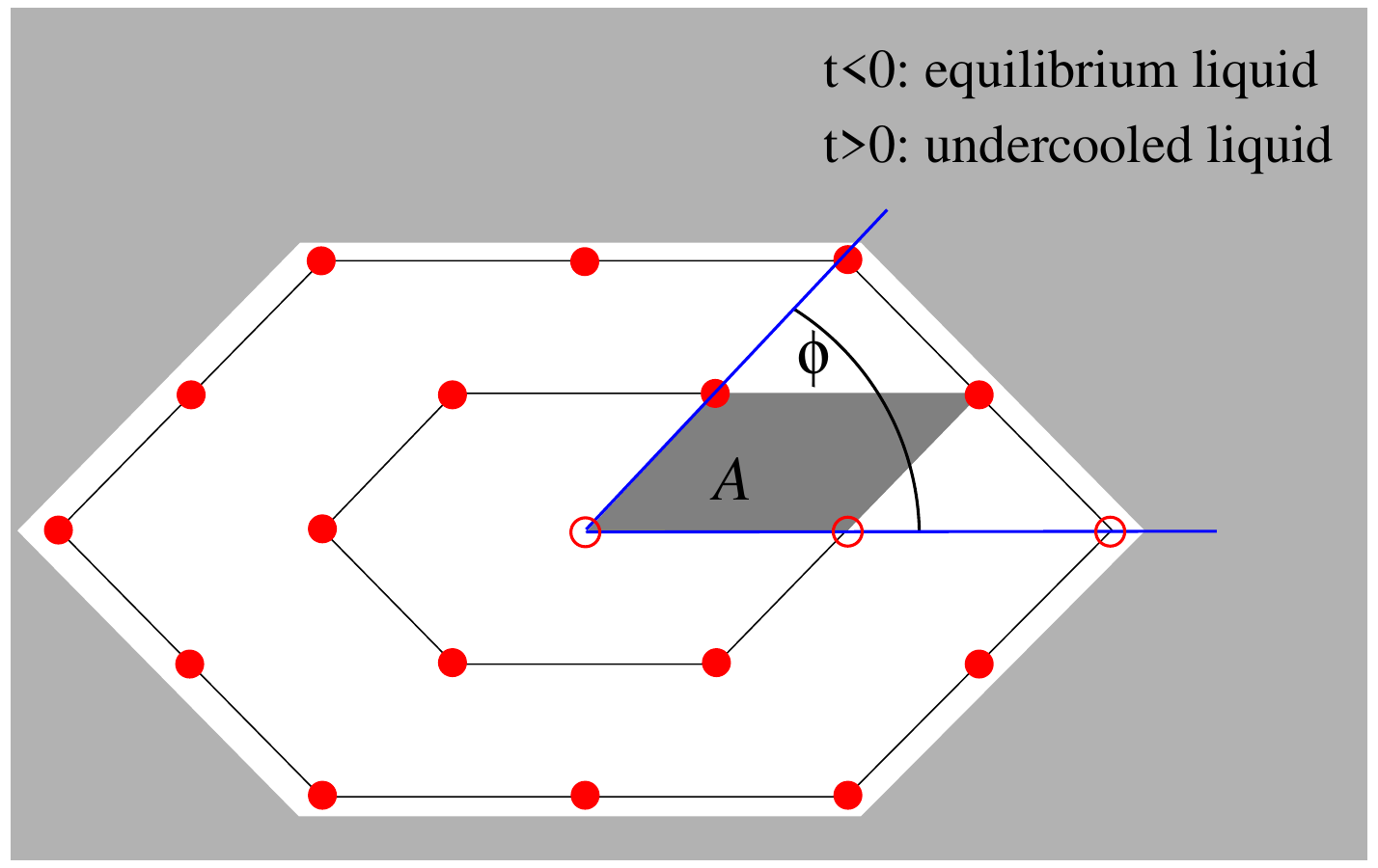}
  \caption{Sketch of the imposed, rhombic nucleation cluster of 19
    particles surrounded by a gray fluid. The angle $\phi$ between the
    spanning basis vectors and the area of a unit cell $A$ are also
    shown.}
  \label{fig:nucleus}
\end{figure}

\begin{figure}
  \includegraphics[width=6cm]{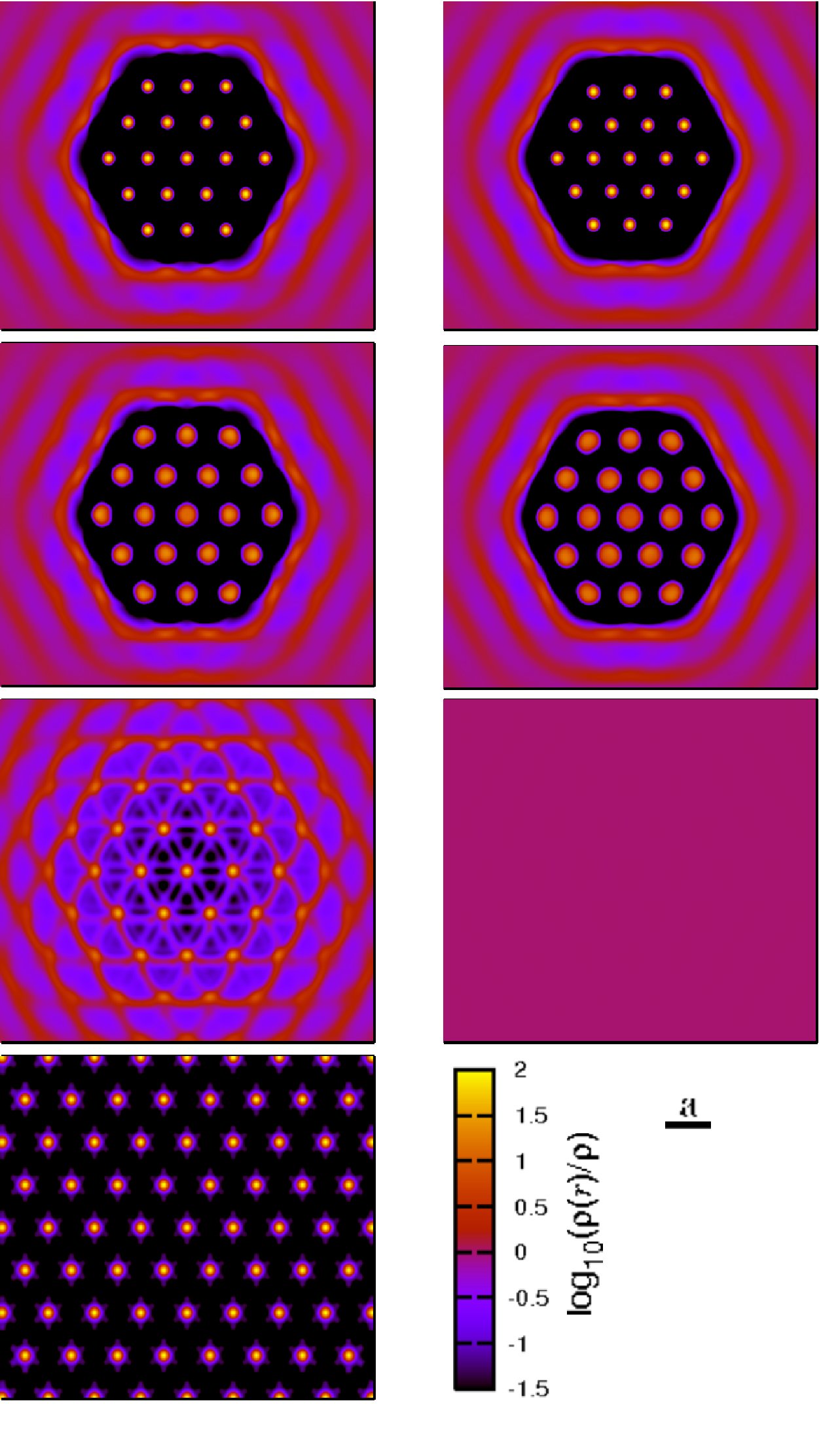}
  \caption{Snapshots of the central region of the dimensionless
    density field $\rho({\bf r},t)/\rho$ of two colloidal clusters
    with strain parameters $A\rho=0.7, \cos\phi=0.5$ (left panel) and
    $A\rho=0.6, \cos\phi=0.5$ (right panel) at times $t/\tau_B=0,
    0.001, 0.1, 1.0$ (from top to bottom; $t/\tau_B=1.0$ only for
    $A\rho=0.7$)~\cite{PRLsupplement}. Note that the images display
    only the system's central region of dimensions $L_x/2\times
    L_y/2$.}
  \label{fig:nucleus_evolution}
\end{figure}

The first setup under study comprises a rhombic nucleation seed of 19
tagged particles, arranged in a hexagon, as sketched in
Fig.~\ref{fig:nucleus}.
The nucleus is characterized by the strain parameters $A$, the area of
a unit cell which in the perfectly ordered hexagonal crystal equals
$A=1/\rho$, and $\phi$, the angle spanned by two of the nucleus axes.
The size of the periodic rectangular box is $16 a \times 16
(\sqrt{3}/2)a$.  In Fig.~\ref{fig:nucleus_evolution} snapshots of the
time-evolving density field are shown exemplarily for two clusters cut
out from two compressed hexagonal crystals with parameters
$(A\rho=0.7,\cos\phi=0.5)$ and $(A\rho=0.6,\cos\phi=0.5)$,
respectively, at times $t/\tau_B=0,0.001,0.1,1.0$.
While the former, less strongly compressed cluster grows into the
equilibrium crystalline state, the latter collapses back into an
undercooled, metastable fluid within $t/\tau_B\lesssim0.1$.  The
growth dynamics of the stable nucleus is a two-stage process: In the
first stage --- on a sub-Brownian time scale $t\lesssim0.002$ --- the
positions of the seed's density peaks move to a cutout of the
thermodynamically stable bulk crystal. In the second stage --- on the
Brownian time scale --- the system crystallizes out of the relaxed
cluster.
\begin{figure}
  \includegraphics[width=6cm]{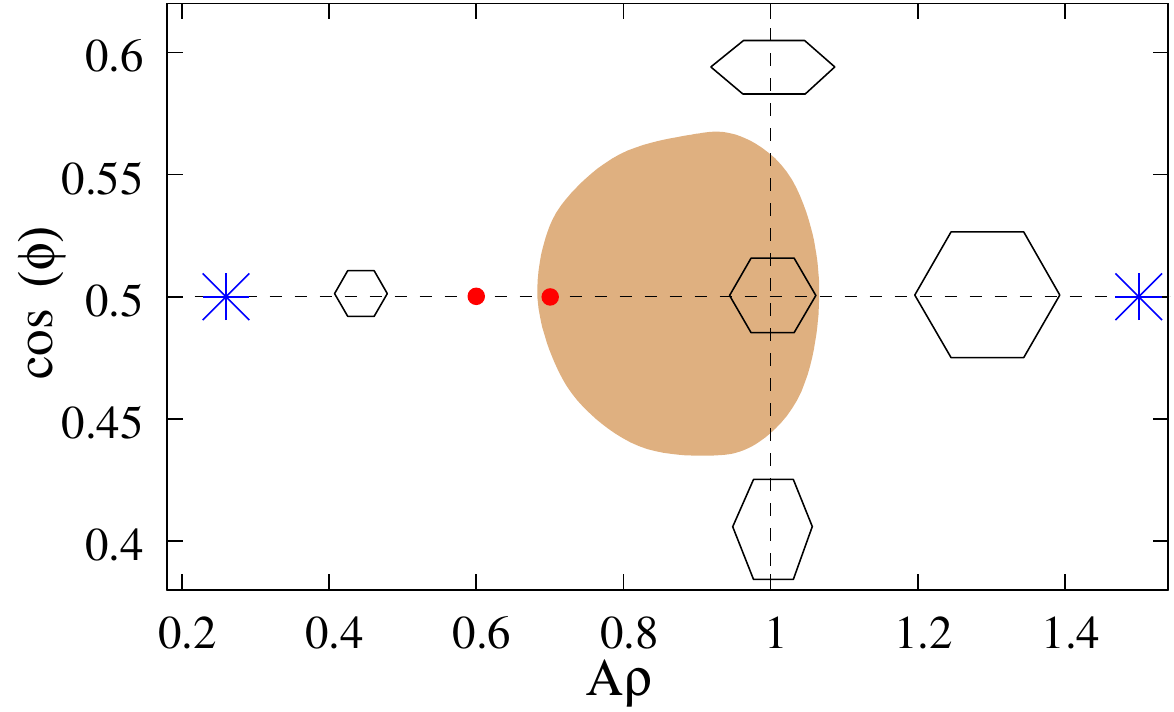}
  \caption{Stability ``island'' of the imposed nucleation cluster of
    19 particles according to Fig.~\ref{fig:nucleus}.  The shaded
    region separates the growth from the no-growth situation. The
    (blue) stars display the according boundaries for fixed
    $\cos\phi=0.5$ obtained from BD computer
    simulation.  The hexagon symbols indicate the way the seeds are
    deformed in the different regions of the parameter space. The
    (red) dots indicate the configurations for the snapshots in
    Fig.~\ref{fig:nucleus_evolution}. }
  \label{fig:width}
\end{figure}

Fig.~\ref{fig:width} displays the ``island'' of growth in the
($A,\cos\phi$)-parameter space, i.e., the set of parameters, for which
the nucleus grows for $t>0$.  It is found that the ``island'' is
nearly symmetric in $\cos\phi$, relative to the equilibrium value of
$\cos\phi=0.5$ while it is asymmetric in unit cell area $A$ about the
ideal value of $A=1/\rho$. This asymmetry is qualitatively validated
by BD simulations~\cite{PRLfootnote_BD,PRLfootnote_BDquench}.

\begin{figure}
  \includegraphics[width=7cm]{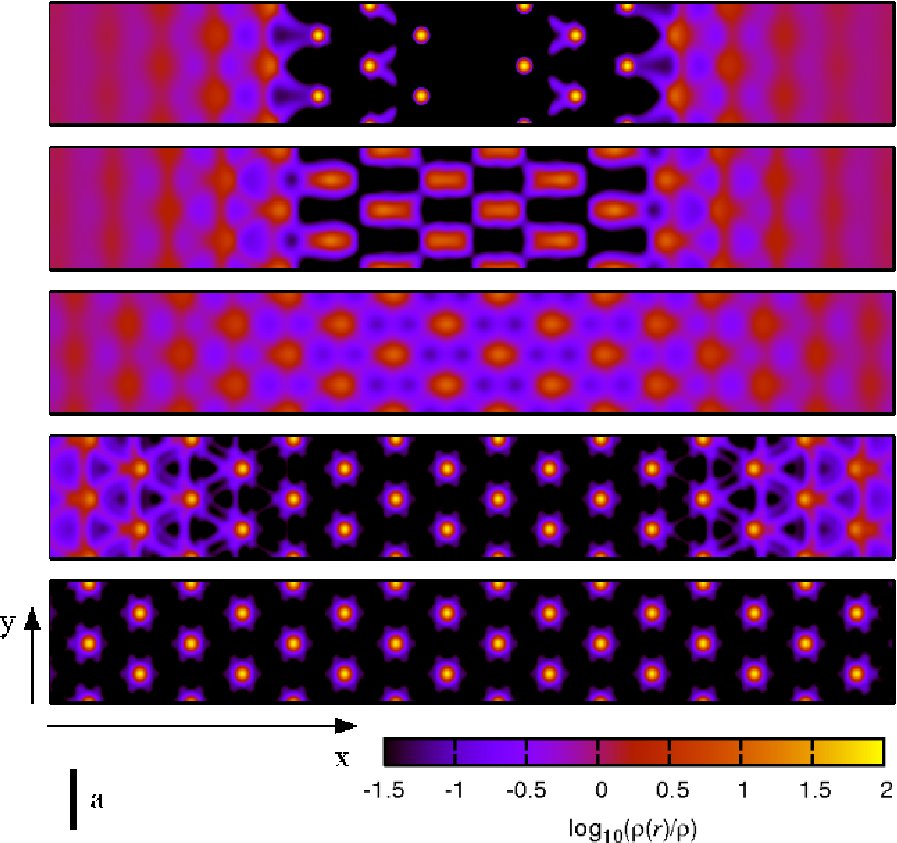}
  \caption{Snapshots of the central region of the dimensionless
    density field $\rho({\bf r},t)/\rho$ of a linear nucleus of two
    times three infinite rows of hexagonally crystalline particles,
    separated by a gap, at times $t/\tau_B=0, 0.01, 0.1, 0.63, 1.0$
    (from top to bottom)~\cite{PRLsupplement}. Note that the images
    display twice the system's central region of dimensions
    $L_x/4\times 2L_y$ for better visibility.}
  \label{fig:array_evolution}
\end{figure}

%ARRAYS*************************************
Within the second setup we study the time evolution of a nucleation
seed of two equal linear arrays along the $y$-direction, each
comprising three infinite rows of hexagonally crystalline particles,
which are separated by a gap, as can be seen from the density map for
$t=0$ in Fig.~\ref{fig:array_evolution}.  These arrays, corresponding
to an equilibrium crystal generated via a suitable external potential,
are displaced relative to each other in $y$-direction by half a
lattice spacing $\Delta y=a/2$. In between the two crystalline arrays
there is an empty stripe of width $\Delta x=\sqrt{3}a$ corresponding
to one missing row of crystalline particles. In contrast to the first
setup, the second setup corresponds to a configuration with a huge
local, non-affine strain relative to a perfect cut-out of a bulk
crystal due to the gap.

In order to keep the gap free of particles during the equilibration of
the surrounding fluid for times $t<0$ we employ an additional strong
external potential in the region of the gap.  The dimensions of the
periodic box within which Eq.~(\ref{eq:ddft}) is solved numerically
are now given by $L_x\times L_y= 64 (\sqrt{3}/2)a \times a$. Snapshots
of the central region of the density field $\rho\left({\bf
    r},t\right)$ are shown in Fig.~\ref{fig:array_evolution} for times
$t/\tau_B=0, 0.01, 0.1, 0.63, 1.0$ after the quench.

Again, a two-stage dynamical scenario is observed: On a sub-Brownian
time-scale of about 0.02 $\tau_B$, the positions of the peaks drift to
those of a perfect cut-out of the stable bulk crystal. This leads to a
rapid filling of the gap. Then crystallization occurs on a Brownian
time-scale.
\begin{figure}
  \includegraphics[width=7cm]{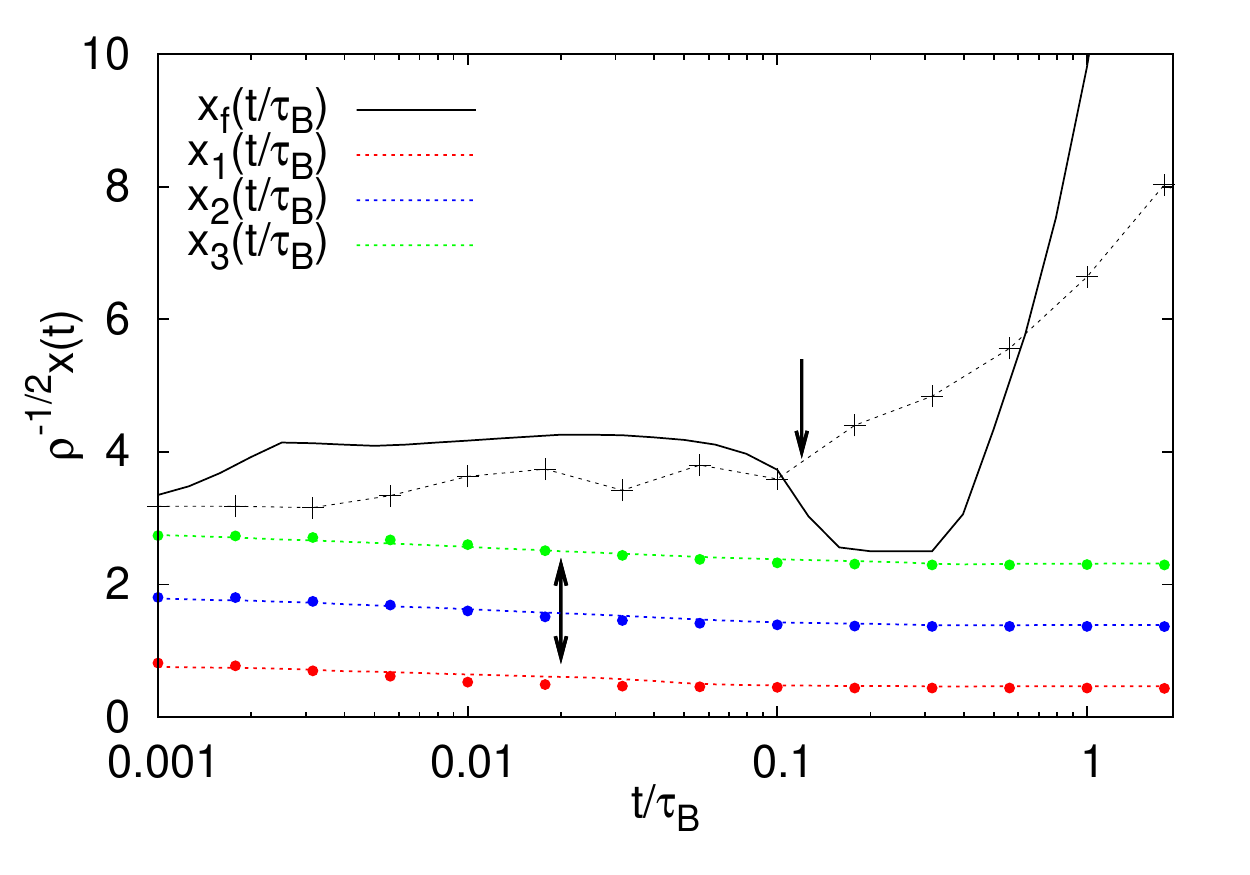}
  \caption{Time evolution of the distance of the linear array's three
    density peaks $x_i(t)$ and of the crystal front $x_f(t)$ with
    respect to the center of the gap as a function of time. Dynamical
    density functional theory results (lines) are compared against
    Brownian dynamics simulation data (symbols; the dashed line
    connecting the crosses is a guide to the eye). The arrows indicate
    the typical time scales on which the relaxation of the $x_i$ is
    occurring and on which the crystal growth sets in, respectively.
  }
  \label{fig:front_row_pos}
\end{figure}
In Fig.~\ref{fig:front_row_pos} we plot the distances $x_i(t)$ of the
three crystalline density peaks and the distance of the crystal front
$x_f(t)$ with respect to the center of the gap as a function of time.
The latter is taken as the inflection point of the envelope function
of the $y$-averaged density field. The theoretical curves are compared
to BD simulation data of the same setup \cite{PRLfootnote_BDquench}
obtained by averaging over the particle positions of $24.000$
independent simulation runs. The two-stage picture is clearly
confirmed.

In conclusion, we have investigated by dynamical density functional
theory whether and how an externally imposed cluster of fixed
particles acts as a nucleation seed for crystal growth if the
particles are released and the system is quenched instantaneously from
a stable to a metastable bulk fluid.  If the imposed cluster is not
too much strained relative to a cut-out of the stable bulk crystal, it
induces global crystallization. The kinetic pathway of the imposed
cluster exhibits a two-stage scenario: the cluster structure first
relaxes towards an appropriate cut-out of the bulk crystal before
further growth. This two-stage process is unexpected since it is
reversed in larger clusters which contain quite a large portion of the
stable bulk crystal. In the latter case crystal growth starts at the
edges but the inner elastic distortion anneals on a much larger time
scale. For higher undercoolings, i.e., larger $\Gamma_>$, the size of
the stability island (Fig.\ 3) increases.

Our predictions can be verified by real-space experiments on
two-dimensional superparamagnetic colloidal particles confined to the
air-water interface in an external magnetic
field~\cite{Zahn:97,Haghgooie:05}.  Qualitatively similar scenarios
are expected for different repulsive interactions and in three spatial
dimensions, which are relevant for nucleation and growth experiments
in sterically and charge stabilized
suspensions~\cite{Vossen:04,Schope:06,Wette:05,deVilleneuve:05}.  In
three dimensions, one may even induce the growth of metastable
crystals and quasi-crystals imposed by suitable nucleation
seeds~\cite{Zander:99}.

\begin{acknowledgments}
  We thank N. Hoffmann, R. Blaak, M. Rex, and A. van Blaaderen for
  helpful discussions. This work has been supported by the DFG through
  the SFB TR6 (project section C3) and the DFT priority program SPP
  1296.
\end{acknowledgments}

%\bibliographystyle{apsrev}
%\bibliography{../../../../bib/journals,../../../../bib/freezing}

\begin{thebibliography}{30}
\bibitem[{PRL({\natexlab{d}})}]{PRLfootnote_BD}
\bibinfo{note}{In the Brownian dynamics simulations, the following cluster
  criterion for crystal growth was chosen. A particle is defined to be
  crystalline if it has six nearest neighbors according to the Voronoi
  construction. Neighboring particles $i$ and $j$ are assigned to the same
  crystalline cluster if $|{\mathrm
  Re}[\Psi_6^i\Psi_6^{j*}]|/|\Psi_6^i\Psi_6^{j*}| \lesssim 0.32$, where
  $\Psi_6^i$ is particle $i$'s complex bond order parameter. Crystal growth is
  defined if the average number of crystalline particles belonging to the
  cluster which contains the seed's innermost particle at $t/\tau_B=0.1$,
  $N_c$, exceeds $N_{\min}=23$. For the strain parameters $\cos\phi=0.5, \rho
  A=0.2,0.4,0.6,0.8,1,1.2,1.4,1.6$ we performed $10.000$ independent runs each
  and measured $N_c$ as a function of $A$. By linear interpolation the
  boundaries of the stability island for $\cos\phi=0.5$ were determined
  $A_{\min}\simeq0.26$, $A_{\max}\simeq1.5$.}
\bibitem[{PRL({\natexlab{e}})}]{PRLfootnote_BDquench}
\bibinfo{note}{Since freezing in the simulation sets in at
  $\Gamma\simeq12$~\cite{Haghgooie:05} we quench the simulated system from
  $\Gamma_<=5$ to $\Gamma_>=20$.}
\end{thebibliography}

\end{document}